\documentclass[printer]{aa}
\usepackage{times}
\usepackage{graphicx}
\usepackage{caption}
\usepackage{longtable}
\usepackage{multirow}
\usepackage{amsmath}
\usepackage{tabu}
\usepackage{longtable}
\usepackage{color}
\usepackage{hyperref}
\usepackage{tabu}

\begin{document}

\title{Selecting asteroids for a targeted spectroscopic survey}

   \authorrunning{Oszkiewicz et al.}
   \titlerunning{Selecting asteroids for a spectroscopic survey}

\author{D.~A.~Oszkiewicz, \inst{1} 
  \and T.~Kwiatkowski \inst{1}
  \and T.~Tomov \inst{2}    
  \and M.~Birlan \inst{3,4}
  \and S. Geier \inst{5}
  \and A.~Penttil\"{a} \inst{6}
  \and M.~Poli\'nska \inst{1}
   }
 
   \institute{Astronomical Observatory Institute, Faculty of Physics, A. Mickiewicz University, 
	      S{\l}oneczna 36, 60-286 Pozna{\'n}, Poland 
         \and Nicholaus Copernicus University, ul. Gagarina 11, 87-100 Toru\'n, Poland
         \and IMCCE - Paris Observatory - UMR 8028 CNRS
	        77 av. Denfert-Rochereau , 75014 Paris, France 
	\and Astronomical Institute of the Romanian Academy,  Strada Cutitul de Argint 5, Bucureti 040557, Romania
        	\and Nordic Optical Telescope, Apartado 474, E-38700 Santa Cruz de La Palma, Spain
	\and Division of Geophysics and Astronomy, Department of Physics, P.O. Box 64, FI-00014 University of Helsinki 
	     }


   \date{Received \today / Accepted xx xx xx} 
   
 \abstract
 {Asteroid spectroscopy reflects surface mineralogy. There are few thousand asteroids whose surfaces have been observed spectrally. Determining the surface properties of those objects is important for many practical and scientific applications, such as for example developing impact deflection strategies or studying history and evolution of the Solar System and planet formation.}
 {The aim of this study is to develop a pre-selection method that can be utilized in searching for asteroids of any taxonomic complex. The method could then be utilized im multiple applications such as searching for the missing V-types or looking for primitive asteroids.}
 {We used the Bayes Naive Classifier combined with observations obtained in the course of the Sloan Digital Sky Survey and the Wide-field Infrared Survey Explorer surveys as well as a database of asteroid phase curves for asteroids with known taxonomic type. Using the new classification method we have selected a number of possible V-type candidates. Some of the candidates were than spectrally observed at the Nordic Optical Telescope and South African Large Telescope.}
{We have developed and tested the new pre-selection method. We found three asteroids in the mid/outer Main Belt that are likely of differentiated type. Near-Infrared are still required to confirm this discovery. Similarly to other studies we found that V-type candidates cluster around the Vesta family and are rare in the mid/oter Main Belt.}
{The new method shows that even largely explored large databases combined together could still be further exploited in for example solving the missing dunite problem.
}


\maketitle

\section{Introduction}

Laboratory analysis of the different groups of meteorites suggested that there must have once been at least 50 to 100 large (150-300 km in diameter) differentiated (into distinct mineralogical layers, that is core, mantle and crust) parent bodies. Those bodies went though a number of disruptive collisions shedding material from their different mineralogy layers and creating hypothetical differentiated asteroid families. Members of those families should have similar orbits and various mineralogical composition across the family. In particular they should contain asteroidal fragments originating from iron core, silicate/basaltic mantle and basaltic crust regions of their parent bodies. 

However up to date Solar System evolution theory, meteoritic evidence and current observations of asteroids are unmatchable. On one side laboratory examination of meteorites indicates existence of multiple parent bodies. Particularly the diverse thermal history (\cite{goldstein1967iron} \cite{kelley20009}) of iron meteorites indicates at least 50 parent bodies cooling at different rates. Analysis of oxygen isotopic ratios (\cite{scott2009oxygen}) in the meteorites from the howardites-eucrites-diogenites (HED) group (commonly believed to be related to the asteroid (4) Vesta) also shows evidence for at least five other than Vesta differentiated parent bodies. However up to date the only presently existing asteroid family that bears evidence of existence of a differentiated parent body is the Vesta family. Present observations of other known asteroid families do not support the hypothesis of multiple parent bodies. Current spectroscopic observations show deficiency of basaltic (originating from mantle and crust) asteroids as compared to the estimations based on meteorite samples (for example \cite{bottkesupplementary} or \cite{burbine1996mantle}).

Furthermore, most of the asteroid families seem to be internally homogeneous. Reflectance spectroscopy of family members shows the homogeneity (\cite{mothe2005reanalysis}) of their surfaces, indicating similar mineralogy within the family. Members of the same asteroid family also often have similar physical properties, such as color indices (\cite{juric2002comparison}) albedos and photometric parameters (\cite{oszkiewicz2011online}). This kind of surface uniformity suggest similar mineralogical composition and homogeneity of the parent body. Asteroid families created during collisions of differentiated parent bodies should contain asteroidal fragments of various composition  (\cite{lazzaro2009basaltic}).

In addition, as mentioned before presently the only asteroid family that is confirmed to originate from a differentiated parent body is the Vesta family.
Vesta family arise from a collisional event that formed a large crater in the surface of asteroid (4) Vesta. Recently NASA's Dawn spacecraft visited Vesta and confirmed the existence of iron core by gravitational field measurements (\cite{russell2012dawn}). Though the presence of iron core was confirmed, the details of Vesta's differentiation are elusive. Two types of Vesta differentiation models are commonly discussed. The magma-ocean models generate eucritic crust overlying diogenite layer, with olivine-rich mantle and a metallic core (for example \cite{Righter1997}). Equivalent model (for example \cite{Barrat2010}) posits that Vesta formed with pockets of magma slowly crystallizing beneath the surface, leading to diogenitic plutons at the crust-mantle boudary or within the basaltic crust. Both of those models were recently challenged by the findings provided by the Dawn mission. The main challenges arrive from the observed amounts and distribution of the olivine material on Vesta inconsistent with the models (\cite{ammannito2013olivine}).

In general, partial differentiation models are also considered possible for some bodies.
 For example high bulk density of Lutetia exceeds density of most common chondric meteorite groups
 (\cite{weiss2012possible}) suggesting that partial differentiation may have led to creation of metallic core overlain by a primitive chondritic crust. This finding is also supported by earlier studies (\cite{carporzen2011magnetic}) which showed an existing remnant magnetic field in carbonaceous chondrites due to core dynamo. Those findings may suggest that other asteroids having chondritic surfaces may in fact be differentiated and having a metallic core  (\cite{weiss2012possible}).

The exact reasons for the discrepancies between the meteoritic evidence and the observational data of asteroids in the Main Belt are yet to be explained. Numerous hypothesis are being tested with the most prevailing being that the differentiated parent bodies were "battered to bits" during collisions and are below our current spectroscopic reach \cite{burbine1996mantle}. Another possible explanation is that space weathering processes (\cite{Chapman2004})could have erased the indicative  spectroscopic features of V-type asteroids and made them look like S-type objects. 
Also shock waves propagated during collisions could hide characteristic spectral features of some materials. For example laboratory spectroscopy of Chelyabinsk meteorites showed that shocked fragments of ordinary chondrites (known to originate from S-type asteroids) can have a spectrum similar to a C-type object (\cite{kohout2013}). Could strong shocking of material mask absorption bands of HED meteorites and their analog V-type asteroids is yet to be tested. Another possibility is that some of the material was somehow selectively removed from the Main Belt - a hypothesis that is yet to be investigated. Evidence of such processes can be seen in other planetary systems. In particular so-called "polluted" spectra of white dwarfs have been found to indicate accreting differentiated objects onto the star (\cite{jura2012two, dufour2010discovery}).  It was also suggested that after collisions fragments of the differentiated parent bodies reaccumulated covering up the basaltic material. Last, but not least our traditional understanding of the differentiation process could be incorrect, perhaps pointing towards the idea of partial differentiation as a more common process in the early Solar System (\cite{weiss2012possible}). Whatever the real explanation is, the search for more evidence for differentiation in the Main Belt is ongoing.

Several surveys and asteroid selection methods targeting the possible remnants of the differentiated material in the asteroid belt, have been published. Most of the selection methods  focus on finding V-type candidate asteroids using the photometric colors obtained by the Sloan Digital Sky Survey (SDSS) (\cite{ivezic2002asteroids}).
For example \cite{roig2006selecting} used the SDSS color photometry in combination with Principal Components Analysis (PCA) to select candidates for a spectral survey of V-type asteroids. The basic idea was to identify the phase-space area where the known V-type asteroids reside and then classify all other asteroids that reside in this area as possible V-type candidates.
\cite{moskovitz2008distribution} presented another method. The method relied on selecting objects having similar photometric colors to those of Vesta family asteroids. Most of the Vesta family asteroids are of V-type, therefore selecting asteroids having similar colors should result in finding other V-type candidates. 
\cite{Carvano2010} derived a new classification scheme for SDSS asteroid colors and investigated the distributions of the new classes in the Main Belt. In particular the Vp class showed a clustering around the Vesta family, and also a scatter in the mid and outer Main Belt.
Another method by \cite{solontoi2012avast} primarily used the $i-z$ color-index to find candidate V-type asteroids. In that method all the asteroids fulfilling the criteria $i-z<- 0.2$ were clasified as possible V-type candidates.

In this study we propose a new selection method that can suggests candidate asteroids for each taxonomic complex. The proposed method assigns spectral complex probability for each asteroid under examination. In contrast to other methods we use multiple databases containing different indicators (here and after features) of spectral complex. In particular we use the Sloan Digital Sky Surveys (SDSS) photometric database, phase function parameter $G_{12}$ database and Wide-field Infrared Survey Explorer (WISE) albedo database. Several asteroids selected by the new method are then targeted using SALT,  and NOT telescopes. Results of those observations are reported. The selection method is compared to similar earlier studies. As opposed to the other methods the technique proposed here can predict objects of any taxonomic complex. This can be especially useful also in applications other than searching for differentiated bodies and their remnants. For example various missions to near earth asteroids look for primitive C-type objects as possible targets. 

In Section 2 we present the novel selection method and possible V-type candidates. In Section 3 we discuss the observations and data reduction. In Section 4 we present our results. Section 5 contains conclusions and discussion.

 \section{Selection method}
To compute probability of an object being of a particular taxonomic complex, we use all the information on asteroid physical properties available up to date in large databases. Previous selection methods focused on using solely the photometric measurements from the Moving Object Catalogue (MOC) obtained in the course of the Sloan Sky Digital Survey (SDSS). In our method we extend the inference by using also other databases, that is
 the database of asteroid phase curves (\cite{oszkiewicz2011online}) and asteroid albedos database from the Wide-Field Infrared Survey Explorer (WISE) (\cite{masiero2011main}). In our method, all those features combined: albedos, slope parameters and SDSS reflectance fluxes determine the probability of an asteroid being of a particular class/complex. 
 
Adding additional features is beneficial for the purposes of preliminary classification in several ways. First, the initial classification can be made for objects which could not be classified based on SDSS solely. For example objects not observed in the course of the SDDS, but observed by the WISE mission and having $G_{12}$ parameter computed. Second, certain features better segregate particular classes than others. For example the C and X types can not be totally segregated given the SDSS data only since the typical differences between the typical SDSS albedos of both classes are within the 10\% error of the data (\cite{roig2006selecting}). On the other hand $G_{12}$ parameter can not distinguish between S and X types due to large distribution overlap, but C types are better segregated from both S and X types (\cite{oszkiewicz2012}). Therefore combing the SDSS data and $G_{12}$ could lead to a more reliable segregation of the C-types.

SDSS data points are available for about 100 000 known objects, albedos for around 100 000 objects and slope parameters for about 500 000 objects. Not all of the data are mutually inclusive, therefore sometimes the probability computation has to be made based on a single feature. This is of-course less reliable than when all the features are available. Some of the data points contain large errors, which also influences the reliability of the resulting probability computation. In contrast to other methods we decided not to remove the data points containing large errors in order not to discriminate possible basaltic candidates. This of course has an impact on the method efficiency, but we consider it a trade-off worth making. The errors can be included in the probability computation. In the subsequent subsections we explain the classification algorithm, the training dataset and uncertainties. To compute probability of an object belonging to specific taxonomic complex, we first estimate its probability to belong to each taxonomic class. The complex probability is then computed by integrating over all the classes in a given complex.

\subsection{Classification algorithm}
In our classification algorithm, we make use of one of the most common classifier in machine learning, that is the Naive Bayes Classifier. In Naive Bayes Classifiers the prediction of a class of an object is made based on the features of the objects which are considered independent of each other (\cite{russell1995artificial}). This means that each of the object features or attributes contributes separately to the identification of a class of an object. Therefore only the variances of the features have to be known, without computing the correlations between the features and the full covariance matrix. Even though the independence assumptions can be sometimes wrong (which they clearly are in case of photometric data, albedo and phase curve parameter), naive Bayes classifiers already proved to be a very efficient tool in many practical applications. One noteworthy advantage of Bayes classifiers is that they can handle missing data by integrating over all possible values of the feature.

In general, the probability of each class can be computed as an a posteriori probability of a class given some specific values of object's features ${\bf{F}} = [F_1,...,F_n]$ for $j= 1,...,n$ (\cite{russell1995artificial}):
\begin{equation}
P(C | F_1, ...., F_n) = \alpha P(C) \prod _{j=1}^n P(F_j | C),
\end{equation}
where:
\begin{itemize}
\item $P(C | F_1, ...., F_n)$ is the a posteriori probability of a class, given object features or attributes $F_1, ...., F_n$,
\item $\alpha$ is the normalization constant,
\item $P(C)$ is the a priori probability of a class and can be estimated as a frequency of specific class (so called informative prior) or assumed uniform among all the classes (so called weakly informative prior),
\item $P(F_j | C)$ is the conditional probability of a feature $F_j$ assuming that the object is of a given class $C$.
\end{itemize}
This can be expanded to:
\begin{eqnarray}
P(C | F_1, ...., F_n) = \alpha P(C) P(P_V | C) P(G_{12} | C) &&\\
P(F_g | C) P(F_r | C) P(F_i | C) P(F_z | C), && \nonumber
\end{eqnarray}
where $P_V$ is WISE albedo, $G_{12}$ is phase curve parameter and $F_g$, $F_r$, $F_i$, $F_z$ are the SDSS albedos (or relative reflectance values). The $F_u$ reflectance flux is omitted from classification - please see section 2.2 for explanation. For example the probability of an object being of a V class could be written as:
\begin{eqnarray}
P( C = V | F) = \alpha P(C =V) P(p_V | C =V) P(g_{12} | C =V) & & \\
P(f_g | C =V) P(f_r | C =V) P(f_i | C =V) P(f_z | C =V), & & \nonumber
\end{eqnarray}
where $p_V$ $g_{12}$ $f_g$ $f_r$ $f_i$ $f_z$ are specific albedo, phase curve parameter and relative flux values of an object and the shapes of probability density functions P are known from templates for a typical V class object. Similarly probabilities for other taxonomic classes can be computed. Probability that an object is of a certain complex rather than of a specific class is computed by integrating over all possible classes in a given complex. For example for S complex the probability could be computed as:
\begin{eqnarray}
P(C = S-complex | F_1, ...., F_n) &  \propto &  \\
\sum_{C_{i=1,..,5} = [S, Sa, Sq, Sr, Sv]}  \prod _{j=1}^n P(F_j | C = c_i). && \nonumber
\end{eqnarray}

In case of the SDSS, WISE and G$_{12}$ data, computing complex probability is more appropriate due to the fact that some of the taxonomic classes are very similar in the visible range. Therefore in this research we focus on computing complex probability.

For the a priori probability $P(C)$ we decided to use uniform distribution to allow the inference to be data driven only. Other a priori distributions (such as for example a frequency prior) could also be used. For example S and C types asteroids are the most common asteroid types, Q and R types are quite rare. To reflect this fact in the inference, one could assume a frequency prior, that is the $P(C)$ would be equal to the ratio of asteroids of a certain type to the whole population. For example for S type asteroids this would be equal to the ratio of number of S type asteroids to asteroids of all classes. Any frequency prior should originate from unbiased distribution of taxonomic classes to reflect the real and not biased frequencies.

The conditional probability $P(F_j | C)$ for a feature $F_j$ can be estimated from the training set of the data as a Gaussian distribution :
\begin{equation}
P(F_j | C) = \frac{1}{\sqrt{2 \pi \sigma_{F_j}}} exp \Bigg{[}-\frac{f_j-\mu_{F_j}}{2 \sigma^2_{F_j}}\Bigg{]},
\end{equation}
where $\mu_{F_i}$ and $\sigma^2_{F_i}$ are feature empirical mean and variance which can be estimated from the training data set or from class templates if known (see the next section for details of obtaining the conditional probabilities). Once all the conditional probabilities  $ P(F_i | C)$ are known, the inference can be made.

Prediction of taxonomic class can be made by computing probabilities for all the Bus-DeMeo classes and selecting the most likely class for an object (\cite{russell1995artificial}):
\begin{equation}
\text{arg max} \Bigg{(} P(C = c_k) \prod _{j=1}^n P(F_j | C = c_k) \Bigg{)}_{ \text{ for } k = 1,..,m}
\end{equation}
The normalization constant $\alpha$ does not have to be known in advance to make a prediction. It can be computed later as $(\sum_{i=1}^{m}P(C_i | F_1, ...., F_6))^{-1}$. The general idea is that objects will be classified based on how well do they fit into the taxonomic templates of reflectance flux, albedo and $G_{12}$s for all the classes. The closer the objects features are to the template of some specific class, the larger the probability that the object is of that class. 

\subsection{Conditional probabilities - training dataset}
 
Naive classifiers belong to so-called supervised learning methods. Those classifiers can be trained based on expert knowledge or training dataset. In our approach we combine both learning techniques. Throughout the study we decided to use the Bus-DeMeo taxonomy which is the most recent commonly used taxonomy. However for some applications (for example classification of the E, M, P classes which can only be separated given albedo) other taxonomies might be more appropriate. The classification scheme described in the previous section is also valid for other taxonomies.

In order to estimate the conditional probabilities of the different classes given albedos we used the WISE albedo database and the JPL Small-Body Database Search Engine combined with Planetary Data System (PDS). From the PDS we have downloaded a database of asteroids used for defining the Bus-DeMeo Taxonomy (EAR-A-VARGBDET-5-BUSDEMEOTAX-1.0). Next we used the JPL tool to extract all the objects with Bus taxonomic class defined. We have translated the Bus taxonomy to Bus-DeMeo taxonomy, using the outline scheme provided in \cite{demeo2009extension}. In cases when the translation was ambiguous we assigned more than one class to the object. For example Ld class in Bus taxonomy corresponds to D or L class in Bus-DeMeo taxonomy, therefore both classes were attributed with a new object. We combined both data sets, the PDS and JPL and then we grouped objects by taxonomic class. Next from the WISE database we have selected objects from our list. In the subsequent step we computed mean WISE albedos and standard deviations for each group of asteroids of the same class. Those parameters are listed in Table \ref{albedo}. The albedo values contributing to the means were weighted by one over their errors. It should be noted that for some taxonomic classes only few asteroids are known. There is no albedo estimates for asteroids from classes O, Sv, Q and Sa, therefore those have to be omitted in our classification. With increasing number of known objects those values can however be refined increasing the overall reliability of the method. Though many of the albedo values for the different classes are overlapping the albedo data still constitute an important piece of information. For example asteroids 14419 and 16352 both have very high albedo above 0.31 suggesting that they might be for example of V-type. The general idea is however not to infer the possible type of the asteroid based on a single parameter, but rather based on a large number of relevant parameters (or features).

\begin{table}
 \caption{Average albedo values for the different taxonomic classes. The columns represent: number of objects (nr), average albedo ($p_V$) and standard deviation ($\sigma_{p_V}$).}
   \label{albedo}
   \centering
   \begin{tabular}{c c c c} 
   \hline\hline  
Taxon & nr & Average $p_V$ & $\sigma_{p_V}$ \\ 
\hline
Cgh & 14 & 0.062 & 0.043 \\
K & 32 & 0.123 & 0.058 \\
A & 11 & 0.216 & 0.078 \\
C & 130 & 0.058 & 0.075 \\
B & 56 & 0.069 & 0.097 \\
D & 26 & 0.09 & 0.155 \\
Sr & 6 & 0.238 & 0.049 \\
Sq & 46 & 0.199 & 0.156 \\
L & 48 & 0.106 & 0.08 \\
S & 370 & 0.21 & 0.088 \\
T & 9 & 0.06 & 0.234 \\
V & 62 & 0.311 & 0.102 \\
X & 93 & 0.071 & 0.149 \\
Ch & 119 & 0.048 & 0.021 \\
Xc & 50 & 0.073 & 0.161 \\
Xe & 20 & 0.14 & 0.202 \\
Cb & 23 & 0.051 & 0.279 \\ \hline
	\end{tabular}
  \end{table}

For the $G_{12}$ we proceed similarly to albedos. That is we extract from the slope parameters database objects with defined taxonomic type and compute mean and standard deviations for each taxonomic class (see Table \ref{gs} for numerical values). There is no G$_{12}$ estimates for asteroids in Sa, Xk, and Sv classes, therefore those also have to be omitted in our classification.

\begin{table}
   \caption{Average G$_{12}$ values for the different taxonomic classes. The columns represent: number of objects (nr), average phase curve parameter (G$_{12}$) and standard deviation ($\sigma_{\text{G}_{12}}$).}
   \label{gs}
   \centering
   \begin{tabular}{c c c c} \hline \hline
Taxon & nr &  G$_{12}$ & $\sigma_{{\text{G}}_{12}}$\\ \hline
Cgh & 20 & 0.615 & 0.228 \\
K & 39 & 0.565 & 0.174 \\
A & 20 & 0.531 & 0.214 \\
C & 153 & 0.643 & 0.169 \\
B & 65 & 0.639 & 0.182 \\
D & 35 & 0.543 & 0.209 \\
Sr & 12 & 0.637 & 0.188 \\
Sq & 100 & 0.425 & 0.186 \\
L & 62 & 1.016 & 0.188 \\
S & 564 & 0.341 & 0.163 \\
T & 18 & 0.331 & 0.126 \\
V & 94 & 0.624 & 0.144 \\
X & 126 & 0.508 & 0.208 \\
Ch & 138 & 0.598 & 0.139 \\
Xc & 70 & 0.368 & 0.223 \\
Xe & 30 & 0.548 & 0.206 \\
Cb & 35 & 0.74 & 0.149 \\
\hline
	\end{tabular}
\end{table}

\onecolumn

\begin{table}[htbp]
 \caption{Template values of the relative reflectance flux for the SDSS filters.}
   \label{var}
   \centering
   \begin{tabular}{c c c c c c c c c c c} \hline \hline
   Type & $F_u$ & $\sigma_{F_u}$ & $F_g$ & $\sigma_{F_g}$ & $F_r$ & $\sigma_{F_r}$ & $F_i$ & $\sigma_{F_r}$ & $F_z$ & $\sigma_{F_z}$ \\ \hline
Cgh &  0.801 & 0.06 & 0.936 & 0.023 & 1.008 & 0.018 & 1.002 & 0.031 & 1.026 & 0.042 \\
Sq &  0.689 & 0.078 & 0.89 & 0.029 & 1.081 & 0.026 & 1.14 & 0.041 & 1.005 & 0.047 \\
A &  0.494 & 0.065 & 0.811 & 0.025 & 1.152 & 0.019 & 1.279 & 0.029 & 1.128 & 0.047 \\
C &  0.927 & 0.045 & 0.976 & 0.016 & 1.009 & 0.012 & 1.014 & 0.019 & 1.004 & 0.026 \\
B &  0.957 & 0.02 & 0.984 & 0.012 & 1.003 & 0.022 & 0.98 & 0.048 & 0.937 & 0.071 \\
D &  0.81 & 0.082 & 0.93 & 0.031 & 1.062 & 0.025 & 1.176 & 0.065 & 1.292 & 0.11 \\
Sr &  0.638 & 0.205 & 0.88 & 0.048 & 1.087 & 0.025 & 1.161 & 0.049 & 1.005 & 0.055 \\
K &  0.766 & 0.053 & 0.916 & 0.021 & 1.063 & 0.023 & 1.127 & 0.041 & 1.075 & 0.045 \\
L &  0.69 & 0.088 & 0.889 & 0.037 & 1.096 & 0.033 & 1.203 & 0.057 & 1.207 & 0.045 \\
S &  0.68 & 0.071 & 0.884 & 0.027 & 1.091 & 0.027 & 1.185 & 0.049 & 1.081 & 0.059 \\
T &  0.824 & 0.042 & 0.939 & 0.013 & 1.052 & 0.008 & 1.147 & 0.014 & 1.209 & 0.017 \\
V &  0.625 & 0.101 & 0.863 & 0.037 & 1.111 & 0.03 & 1.168 & 0.064 & 0.766 & 0.101 \\
X &  0.898 & 0.051 & 0.967 & 0.019 & 1.022 & 0.007 & 1.076 & 0.019 & 1.118 & 0.033 \\
Ch &  0.911 & 0.041 & 0.975 & 0.014 & 0.988 & 0.009 & 0.981 & 0.018 & 0.994 & 0.029 \\
Cb &  1.019 & 0.03 & 1.006 & 0.011 & 1.0 & 0.007 & 1.017 & 0.006 & 1.025 & 0.031 \\
Xe &  0.843 & 0.03 & 0.931 & 0.033 & 1.067 & 0.032 & 1.116 & 0.037 & 1.124 & 0.049 \\
Xc &  0.868 & 0.067 & 0.956 & 0.018 & 1.029 & 0.006 & 1.074 & 0.007 & 1.096 & 0.005 \\ \hline
	\end{tabular}
  \end{table}

\twocolumn

\begin{figure}[htbp]
   \centering
   \resizebox{\hsize}{!}{\includegraphics{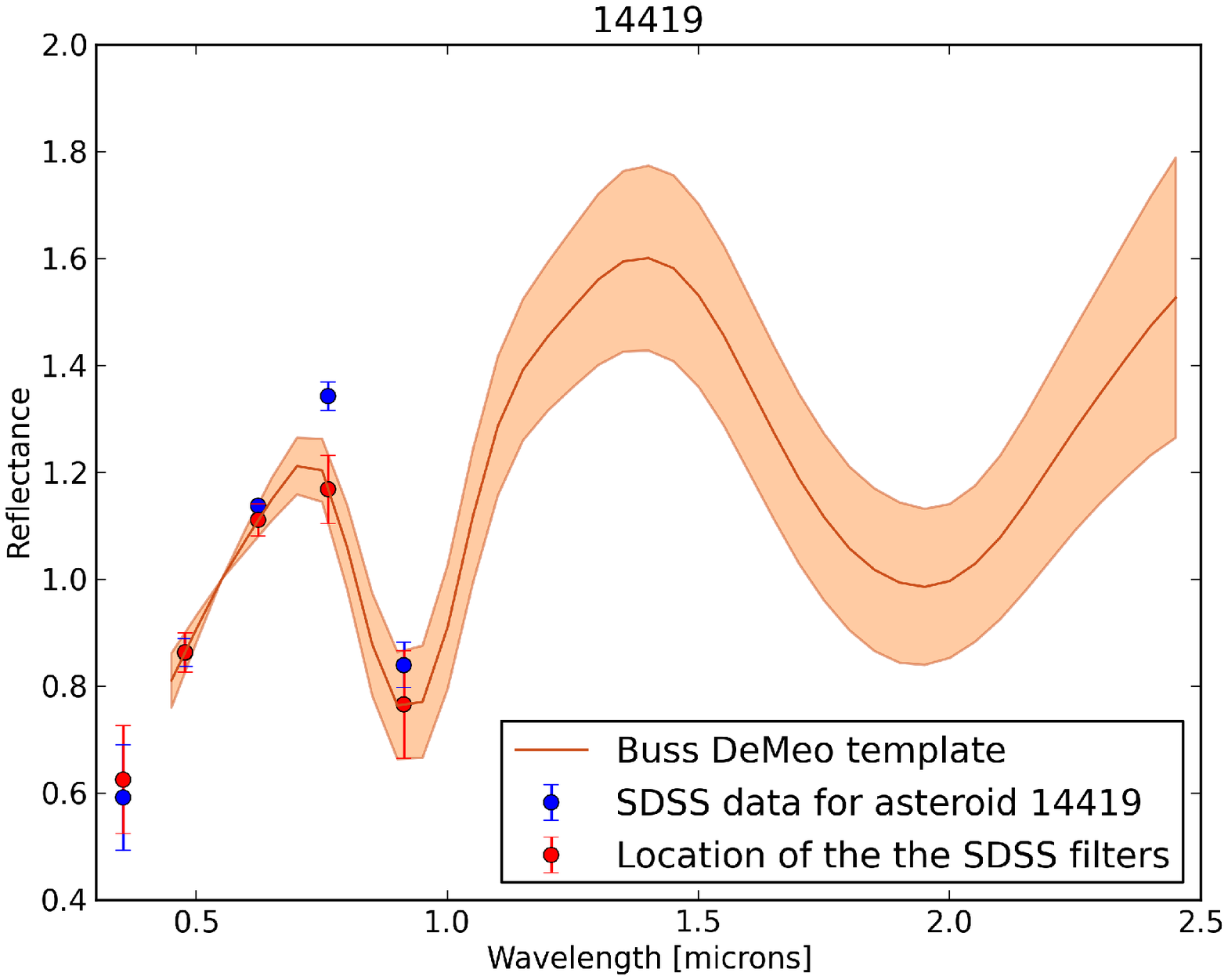}}
   \caption{SDSS data points for asteroid (14419) 1991 RK23 as compared to the Bus-DeMeo V-class template. The shaded area denotes standard deviation of the template. The u filter reflectance is added as an linearly extrapolated point for reference only. }
   \label{14419}
\end{figure}

\begin{figure}[htbp]
   \centering
   \resizebox{\hsize}{!}{\includegraphics{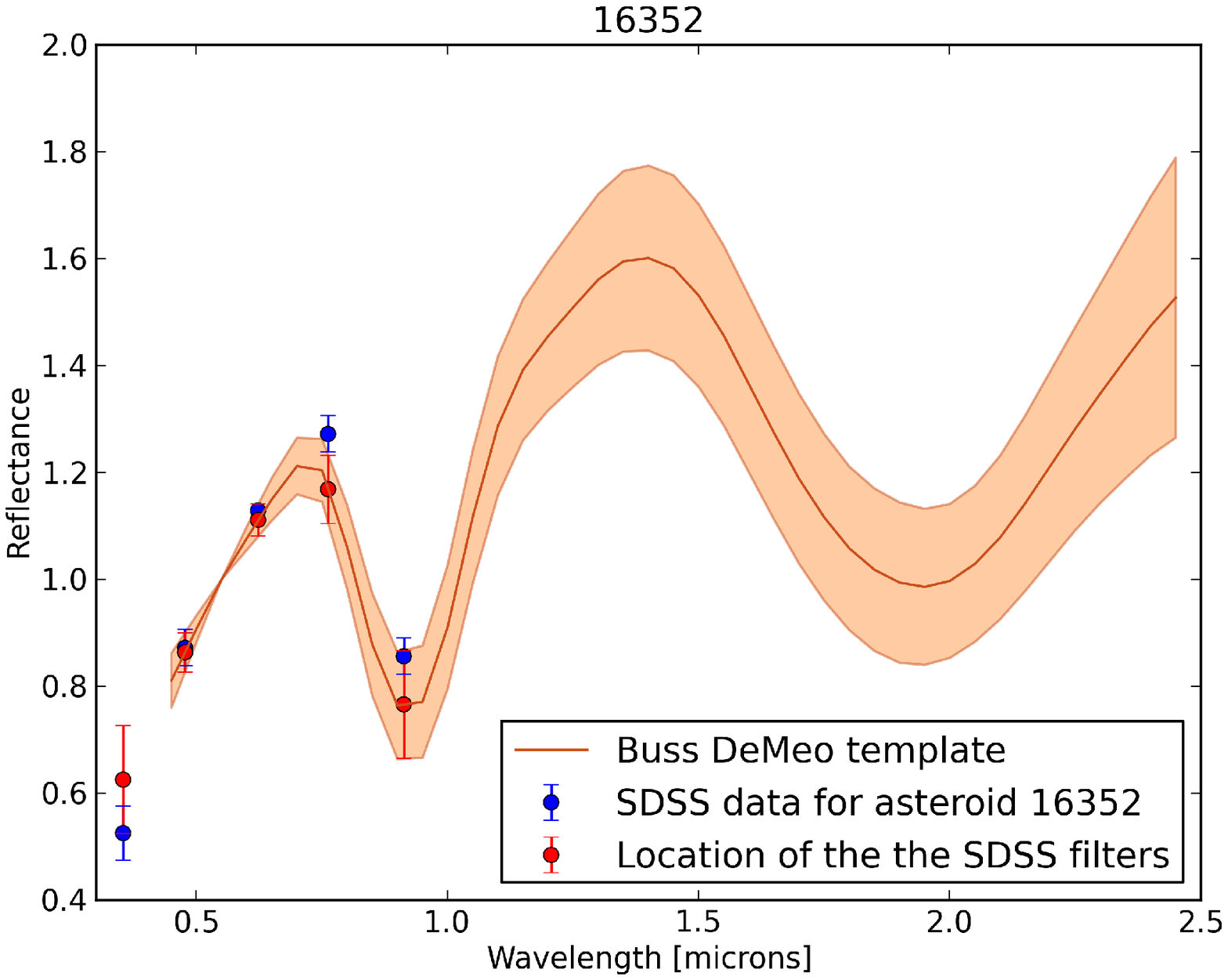}} 
   \caption{SDSS data points for asteroid (16352) 1974 FF as compared to the Bus-DeMeo V-class template. The shaded area denotes standard deviation of the template. The u filter reflectance is added as an linearly extrapolated point for reference only. }
   \label{16352}
\end{figure}

\begin{figure}[htbp]
   \centering
   \resizebox{\hsize}{!}{\includegraphics{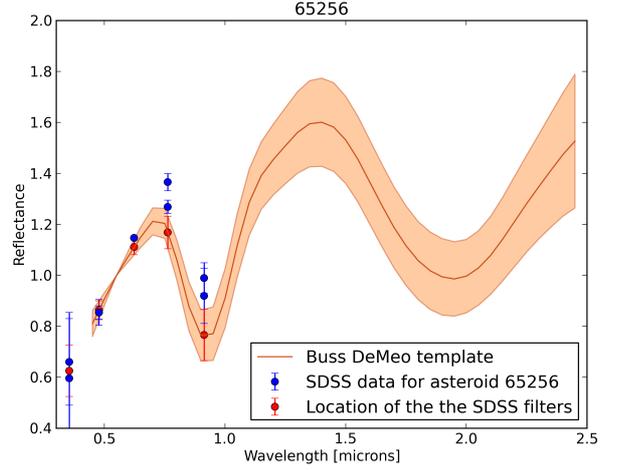}} 
   \caption{SDSS data points for asteroid (65256) 2002 FP34 as compared to the Bus-DeMeo V-class template. The shaded area denotes standard deviation of the template. The u filter reflectance is added as an linearly extrapolated point for reference only. }
   \label{65256a}
\end{figure}



For reflectance fluxes $\mu_{F_j}$ and $\sigma_{F_j}$  can be estimated from the spectroscopic templates. Those templates are for example available for Bus-DeMeo taxonomic classes via the MIT website. We present the means and standard deviations of reflectance fluxes for the SDSS filters per taxonomic class estimated based on Bus-DeuncertaintyMeo taxonomic templates in Table \ref{var}. The taxonomic templates list mean reflectance values and their standard deviations per taxonomic class every 0.05 microns. The centers of the SDSS filers bands ($u = 0.3543$~$\mu m$, $g =0.4770$~$\mu m$, $r= 0.6231$~$\mu m$, $i=0.7625$~$\mu m$, $z=0. 9134$~$\mu m$) lie between the data points provided in the templates, therefore the mean reflectance values for the SDSS filters have to computed by linear interpolation between the adjacent template points. The corresponding standard deviations for the mean reflectance values in the SDSS filters are also computed accordingly. The u filter reflectance can be computed by a linear extrapolation. However several of the taxonomic classes show a drop in reflectance in the UV range, and therefore such extrapolation is unreliable and could introduce additional errors. Furthermore, since we are trying to match the data to the Bus-DeMeo classes (which do not extend as far into UV range) including the u filter would just introduce another source of uncertainty. The u filter reflectance values are therefore not used in the classification process. In table \ref{var} we list the extrapolated reflectance values for the u filter for a refference only.

It should be noted that the taxonomic templates for types O, R, and Cg have very large error envelopes ($\sigma$ values -0.999 as listed on the MIT webpage). Because the Cg, O, and R classes were deÞned by a single object the standard deviation is set to -0.999. Therefore those classes are omitted from classification altogether.

Figures \ref{14419}, \ref{16352}, \ref{65256a} illustrate the Bus-DeMeo template (shaded area represents the standard deviation for the class) for a V-type asteroid and the interpolated points for the SDSS filters. Any asteroid having SDSS reflectance values fitting the corresponding range of the intra/extrapolated SDSS points (denoted in red in the Figure) will have a high V-complex probability and therefore will be classified as a potential V-type candidate. Similar plots can also be made for other taxonomic classes. In these figures we also plot the computed reflectance values for three selected asteroids from the SDSS database. Asteroid 65256 is an outer Main Belt object and it was observed by the SDSS twice. Asteroids 14419 and 1652 are inner Main Belt asteroids, each observed once by the SDSS. 

To compare the SDSS against the spectroscopic templates it is necessary to compute the reflectance values from the raw SDSS photometric data points. To perform this task we adopt the method by \cite{roig2006selecting}. First the SDSS color indices have to be corrected by solar colors:
$c_{u-r} = (u-r) -1.77$, $c_{g-r} = (g-r)-0.45$, $c_{r-i} = (r-i) -0.10$, $c_{r-z}=(r-z)-0.14$. The albedos at each band center, normalized to the albedo at the r band, are then : $F_u = 10^{-0.4c_{u-r}}$, $F_g = 10^{-0.4 c_{g-r}}$, $F_i=10^{0.4 c_{r-i}}$, $F_z=10^{0.4 c_{r-z}}$. The errors can be estimated as: $\Delta F / F = 0.9219 \Delta c (1+0.4605 \Delta c)$, where $\Delta c$ are color errors computed as the usual root squared sum of  the corresponding magnitude errors. For more details please see \cite{roig2006selecting}. The computed flux values have to be then slide up/down to the template. This is done by aligning the 0.55 microns normalization point of the template and the data.

Two things should be noted: i) given a different set of asteroids in each taxonomic complex can result in slightly different means and variances for albedo and G12 parameter; ii) with increasing number of taxonomic classifications those parameters will be better refined and also correlations between the different features will be better known.

\subsection{Including uncertainties}
For better probability estimation there is an option of including the features uncertainties in the probability computation. 

Each measurement is fraud with both systematic and random errors ( for example some of the SDSS measurements show large photometric errors). Therefore the real value of a feature can lie anywhere in the given uncertainty region. Each feature can therefore be represented by a large number of possible values. Each of the features is sampled within its uncertainty. Next classification as described in the section above is performed. This results in a large number of classifications (and probabilities $p(c\mid f)$ for a specific class $c$ and specific sampled set of features $f$) per single object. Objects having small feature uncertainties will often be categorized to the the same class. Objects having large feature uncertainties will be categorized to many different classes. The final decision classifying the object can be made based on how many times the object was classified into what category.  The final probability for a given class C including the uncertainties can then be computed as:

\begin{equation}
P(C) = \int_{F_L} ^{F_U} \! p(C\mid \bf{F}) \, \mathrm{d}\bf{F}.
\end{equation}
Where {\bf{F}} is a set of features and $F_U$, $F_L$ are the upper and lower boundary values obtained from features uncertainties. In practice the taxonomic class probability can be computed as a weighted fraction of class classifications as compared to all classes:
\begin{equation}
P(C) = \frac{\sum_{j=1}^{N_C} w_j}{\sum_{k=1}^{N_{all}} w_{k}},
\end{equation}
where $N_C$ is the number of times an object was classified as a given taxonomic class and $N_{all}$ is total number of classifications. The $w_j$ and $w_k$ weights are the probability values $p(c\mid f)$ computed for the class under cosideration and all the other classes respectively.

\subsection{Comparison with other methods and validation}

To compare our method with those already available in literature, we perform classification of basaltic candidates indicated in   \cite{moskovitz2008distribution}, \cite{roig2006selecting} and \cite{solontoi2012avast}. In particular we compare the classification for all the numbered asteroids in the candidate lists that have all the features (colors, albedo, $G_{12}$s)  available. Additionally we check how many of those objects were observed spectrally and confirmed V-type. In Table \ref{comparison} we list the classification and probabilities for those objects.

Out of  50 candidates listed by \cite{moskovitz2008distribution}, 9 numbered objects have all the features available.  Out of those all, but two ((55092) 2001 QO123, (50802) 2000 FH27) were classified as V-types candidates using our method. (50802) 2000 FH27 has been observed twice by the SDSS and the two datasets give completely different classifications (96 \% D-type or 57\% Sr-type). (55092) 2001 QO123 produces a better fit to Sr-type in our method. It is worth to notice that those objects were also not identified by \cite{roig2006selecting} as V-type candidates. Spectroscopic observations and classification is available for 7 objects from this list. Asteroid (46690) 1997 AN23 was classified as S-type object after spectroscopic observations, (\cite{moskovitz2008distribution}) even though it was listed as V-type candidate by all the authors. The object also had 72\% probability of V-type in our method. It is worth to notice that the object was observed by the SDSS four times. The classification based on the other observations in our method leads to S-complex (Sr type) classification three out of four times (with probabilities of 52\%, 52\%, 64\%). Spectroscopic observations are also available for 6 other objects from this list. All of those were classified as V-type based on spectra. All of those were also classified as V-type candidates in our method. In general our method is in good agreement with that of \cite{moskovitz2008distribution} (77\% coverage among the candidates) and spectroscopic observations for those objects.

Out of 233 V-types candidates that are members of the Vesta family indicated by \cite{roig2006selecting} there are 71 numbered asteroids that have all the features available. Out of those all but three objects ((20437) 1999 JH1, (10157) Asagiri, (16452) Goldfinger) were classified as V-types candidates using our method. Out of  266 V-types candidates that are not members of the Vesta family indicated by \cite{roig2006selecting}, 78 numbered objects have all the features available. Out of those all, but two ((44496) 1998 XM5, (10666) Feldberg) were also classified as V-type using our method. The objects having non-matching classification are mostly classified as Sq or Sr in our method. The S-complex classifications agree between the different SDSS observation sets. In general  144 objects (out of 149) indicated by  \cite{roig2006selecting} were also classified as V-type candidates in our method. Out of those objects 17 were spectrally observed and confirmed V-type. All of those 17 objects were classified as a V-type candidates by our method. There is a good agreement of our method with that of \cite{roig2006selecting} (96\% candidate coverage among the non-vestoids and vestoids) and spectroscopic observations.


Out of 2023 numbered objects fitting the AVAST selection criteria ($i-z<-0.2$) 673 have all the features available and 25 have been observed  spectrally \cite{solontoi2012avast}. All of the observed objects turned out to be V-types except for 46262 (S-type) and 27202 (A-type). Two other objects were missclassified by our method, namely 20455 and 32272 for which the spectroscopic observations are indicative of V-type, but the classification places them to be S-complex objects. Overall our method predicted the right complex for 21 out of 25 objects observed and is in good agreement with the spectroscopic observations. However large number of V-type candidates fitting the AVAST criteria is classified as S- C- or X- complex objects in our method. Though it should be noted that only two object from the non-matching candidate list (32272 and 20455) were observed.

Overall, the method classified correctly most of the spectrally confirmed objects and has a good agreement in candidate selection with the methods by \cite{moskovitz2008distribution} and \cite{roig2006selecting}. There is less percentage of overlap with the basaltic candidates indicated by \cite{solontoi2012avast} though. However our method predicted the taxonomic complex correctly for 88\% of the objects observed listed in table \ref{comparison}.

\subsection{V-type candidates}

\begin{figure}[htbp]
   \centering
   \resizebox{\hsize}{!}{\includegraphics{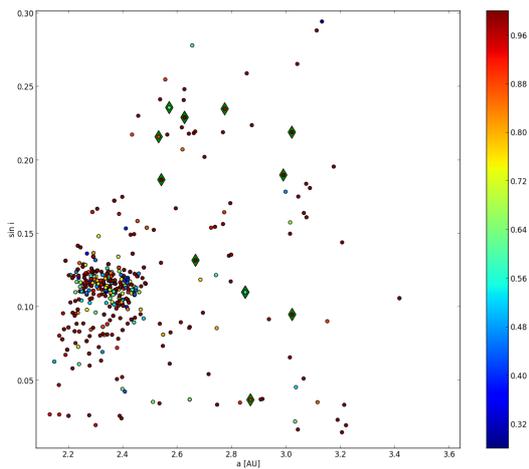}} 
   \caption{Distribution of V-type candidates in the Main Belt. Color corresponds to probability. Candidates with good quality data are additionaly denoted with a diamont marker.}
   \label{dist}
\end{figure}

We have predicted taxonomic complexes for all the asteroids present in the given datasets. In here we will examine asteroids for which all the features (SDSS photometry, WISE albedo and G$_{12}$) are available. In Fig. \ref{dist} we present the distribution of V-type candidates as predicted by our method. Most of the V-type candidates are clustered around the Vesta family, but also scattered in the mid and outer Main Belt. Those are objects which were classified as V-type candidates at least once (based on one set of observations). Other than the Vesta family there seems to be no clustering of candidates around other families. In table \ref{candi} we list our candidates located in mid and outer Main Belt. Though the list is extensive, most of the candidates have SDSS data uncertainties $> 10$ \% or classifications based on different datasets disagree with each other - therefore the classification should be taken with caution. Our top candidates (classification based on data with smaller uncertainties) include 11 first asteroids listed in table \ref{candi}. Out of those (21238) Panarea was already confirmed as a V-type asteroid (\cite{Binzel2006}, \cite{Hammergren2006}, \cite{roig2008v}). (10537) 1991 RY16 is an R-type object (also originating from a differentiated body) previously observed by \cite{Moskovitz}. 40521 (1999 RL95) was previously indicated as basaltic candidate and also observed \cite{roig2008v}.

\section{Observations and data reduction}

We have selected a number of objects to be observed. The objects are listed in Table \ref{sel}. We focus on objects that are located in the mid and outer Main Belt as well as objects located in the inner Main Belt, but outside the Vesta dynamical family (Definition of Vesta family extracted from HCM Asteroid Families V2.0. EAR-A-VARGBDET-5-NESVORNYFAM-V2.0. NASA Planetary Data System, 2012). We have avoided candidates that were targeted by other V-type candidate surveys.

\onecolumn

\begin{table}
\caption{Asteroids selected for observations.} 
   \label{sel} 
   \centering
   \begin{tabular}{c c c c c c} 
   	\hline \hline
	Asteroid  & Location & albedo & i-z & G$_{12}$ & prob\\ 
	 number  &  in MB &  &   &   \\ \hline 
	
	15551 & outer & 0.28  $^{\pm 0.03}$ & -0.15  $^{\pm 0.04}$ & 0.50  $^{+0.16}_{-0.16}$ & S 0.72 ( 0.01 ) \\ 
	65256  & outer  & & -0.23 $^{\pm 0.08}$ & 0.80  $^{+0.36 }_{-0.40}$ & A 0.79 (0.1)\\
	1979 & inner  & 0.39   $^{\pm 0.04}$ & -0.21  $^{\pm0.04}$ & 0.02  $^{+0.13 }_{-0.11}$ & Sr 0.43 (0.04)\\
	10484   & inner  & 0.23   $^{\pm 0.06}$  & -0.23  $^{\pm0.03}$ & 0.51  $^{+0.13 }_{-0.13}$ & Sr 0.43 (0.18)\\
	41485   & inner  & & -0.44  $^{\pm 0.03}$ & 0.47  $^{+0.17 }_{-0.17} $ & V 0.67 \\
	13679   & inner  & 0.31   $^{\pm 0.03}$ & -0.12  $^{\pm 0.03}$ & 0.61  $^{+0.15 }_{-0.15}$ &  A 0.59 ( 0.01 ) \\
	18775  & inner  & & -0.17  $^{\pm 0.05}$ & 0.74  $^{+0.13 }_{-0.14}$ & A 0.37 (0.26)\\ 
	30243  & inner  & & -0.25  $^{\pm 0.06}$ & 0.50  $^{+0.19 }_{-0.20} $ & V 0.49 \\ 
	40521   & mid  & 0.28   $^{\pm 0.03}$ & -0.32  $^{\pm 0.07}$ & 0.39  $^{+0.17 }_{-0.17}$ & V 0.89 ( 0.89 ) \\
	31455   & mid  & & -0.21  $^{\pm 0.07}$ & 0.50  $^{+0.16 }_{-0.16}$ & L 0.38 (0.26)\\
	15717  & mid  & & -0.07  $^{\pm 0.16}$ & 0.60  $^{+0.12 }_{-0.12}$ & V 0.49 \\
	33493  & mid  & & -0.13  $^{\pm 0.04}$ & 0.44  $^{+0.12 }_{-0.12}$ & L 0.41 (0.16)\\ 
	11699   & inner  & 0.23   $^{\pm 0.06}$ & -0.28  $^{\pm 0.04}$ & 0.36  $^{+0.11 }_{-0.11}$ & Sr 0.31 (0.18)\\ 
	14419   & inner  & 0.32   $^{\pm 0.04}$ & -0.47  $^{\pm 0.06}$ & 0.45  $^{+0.12 }_{-0.12}$ & V 1.0\\	
	16352  & inner  & 0.37   $^{\pm 0.04}$ & -0.39  $^{\pm 0.06}$ & 0.51  $^{+0.13 }_{-0.13}$ & V 0.99 ( 0.99 ) \\ 
	\hline
	
   \end{tabular}

\end{table}
\twocolumn

The observations were performed in 2012 at two different telescopes: the Nordic Optical Telescope (NOT), and the South African Large Telescope (SALT). NOT is located in the Canary Islands and SALT is located in South Africa. 

At the 2.56 m NOT we used the Andalucia Faint Object Spectrograph and Camera (ALFOSC) combined with low resolution grism number 11 and slit 1.8'' for all of our targets except for one. For asteroid (15551) Paddock we used grism 12 with 2.5'' slit as the object was observed during a test night for which Grism 11 was not available. Grism 11 gives wavelength coverage from 0.39 microns to 0.995 microns and has a dispersion of 4.8 {\AA} per pixel. Grism 12 gives wavelength coverage 0.51 microns to 1.1 microns and has a dispersion of 13{\AA} per pixel. Grism 12 has a build-in blocking filter OG515. No other blocking filters were used. Differential tracing was used. 

%

At the 10 m SALT telescope (\cite{Buckley+06})
we used the Robert Stobie Spectrograph (RSS; \cite{Burgh+03}, \cite{Kobulnicky+03}). The $4\arcsec$ slit was imaged through the
pc04600 order blocking filter onto the pg0300 grating, which resulted in the
FWHM resolution $\Delta \lambda=46$\AA ($R\simeq 150$).  The spectrum was
recorded with the mosaic CCD camera in the useful range from 4660 {\AA} to
9000 {\AA}, with a gap (due to a physical gap between the three CCD segments) from 7746{\AA} to 7917{\AA}.  In this setup the reciprocal
dispersion at the center of the spectrum was 1.5{\AA} per pixel, therefore we used the 4x4 bining, which gave us 7 superpixels per FWHM resolution.

At the time of observations the non-sidereal tracking of the telescope was
not commissioned yet, so we oriented the spectrograph slit along the
asteroid trail and took 16 consecutive exposures of 120~s.
During the
exposure time, sky movement of each of the observed asteroids was equal to the seeing, the FWHM of which was $1\farcs2$ -- $1\farcs4$.  
This procedure worked as a
natural dithering, which helped to average-out the fringing pattern at the
red end of the spectra.  It should be noted that thanks to the SALT
Atmospheric Dispersion Compensator and a wide slit the influence of the
differential refraction was minimized.  For each asteroid a solar analog
star was observed at a similar airmass.

Primary reduction of the data was done on-site with the SALT science
pipeline (\cite{Crawford+10}).  After that, the bias and gain corrected and
mosaicked long-slit spectra were reduced in a standard way with the IRAF
package.  The spectrum of Ar lamp was used to calibrate the wavelength
scale as well as spectral flats to correct for pixel-to-pixel variations. After median-combining all individual spectra of a given asteroid, the
obtained spectrum was divided by the spectrum of the solar analog star.  


\section{Results}

\begin{figure}[htbp]
   \centering
   \resizebox{\hsize}{!}{\includegraphics{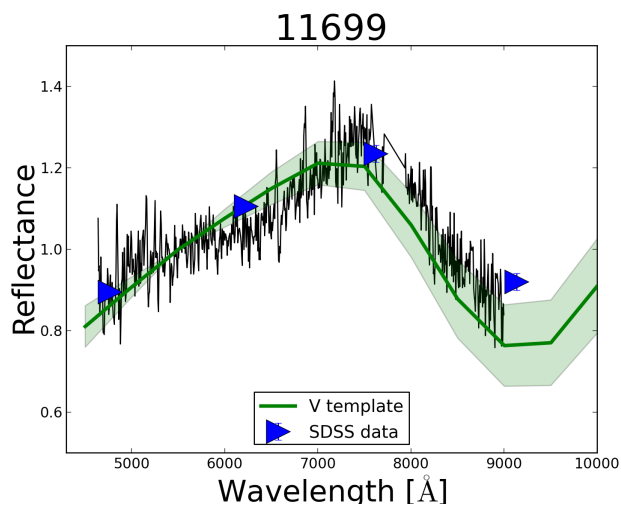}} 
   \caption{Spectra of asteroid (11699) 1998 FL105 obtained at the SALT. The triangles denote the reflectance values computed from the SDSS photometry. The thick line represents the taxonomical template for the V-type and the shaded area is the standard deviation of the template.}
   \label{spec1}
\end{figure}

\begin{figure}[htbp]
   \centering
   \resizebox{\hsize}{!}{\includegraphics{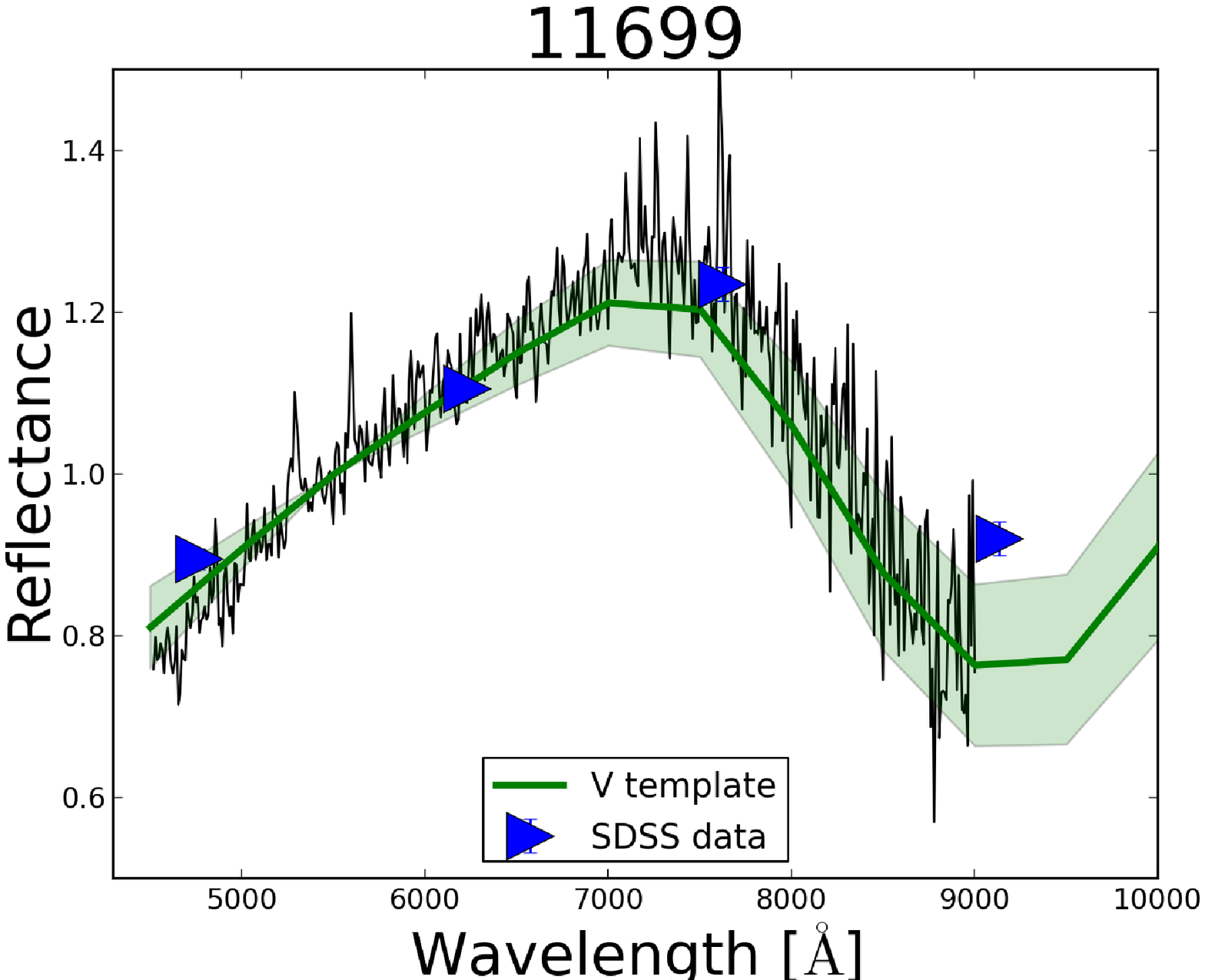}}
      \caption{Spectra of asteroid (11699) 1998 FL105 obtained at the NOT. The triangles denote the reflectance values computed from the SDSS photometry. The thick line represents the taxonomical template for the V-type and the shaded area is the standard deviation of the template.}
      \label{spec2}
\end{figure}

\begin{figure}[htbp]
   \centering
   \resizebox{\hsize}{!}{\includegraphics{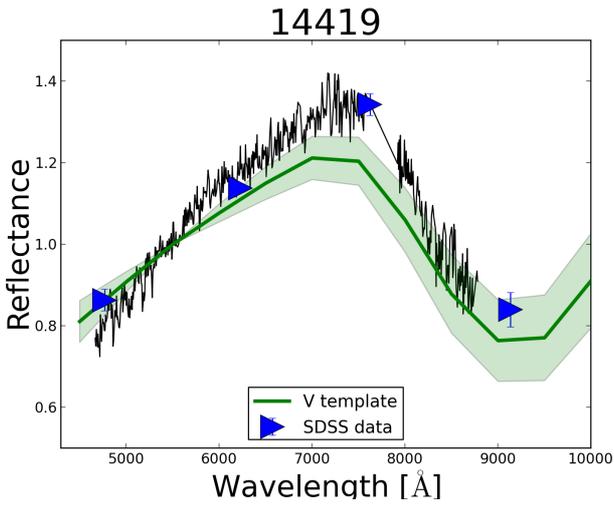}}
      \caption{Same as in Fig. \ref{spec1} for asteroid (14419) 1991 RK23.}
      \label{spec3}
 \end{figure}

\begin{figure}[htbp]
   \centering
   \resizebox{\hsize}{!}{\includegraphics{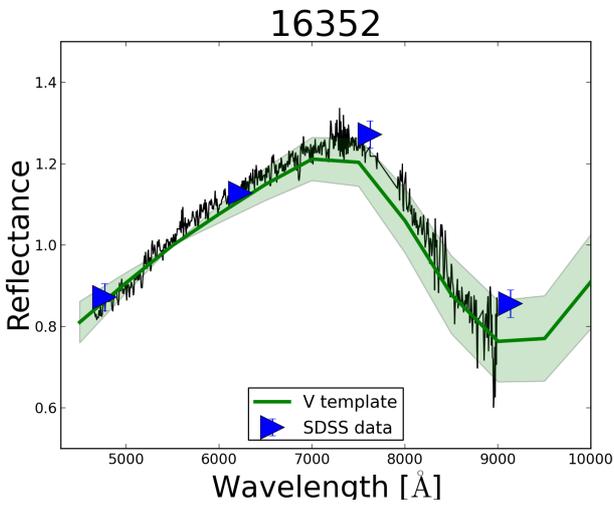}}
   \caption{Same as in Fig. \ref{spec1} for (16352) 1974 FF.}
      \label{spec4}
 \end{figure}

\begin{figure}[htbp]
   \centering
  \resizebox{\hsize}{!}{\includegraphics{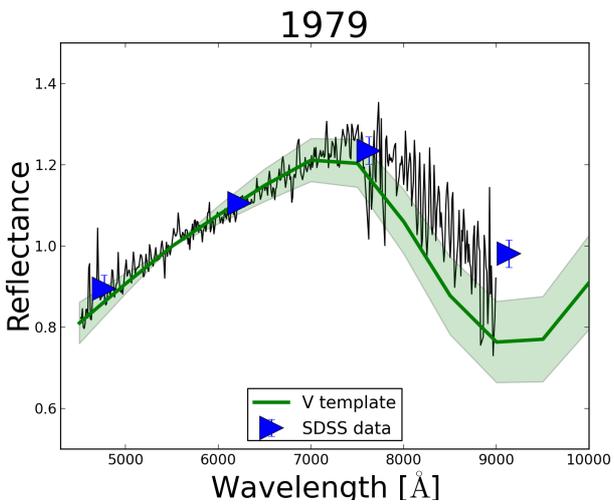} }
  \caption{Same as in Fig. \ref{spec2} for asteroid (1979) Sakharov.}
     \label{spec5}
 \end{figure}

\begin{figure}[htbp]
   \centering
\resizebox{\hsize}{!}{\includegraphics{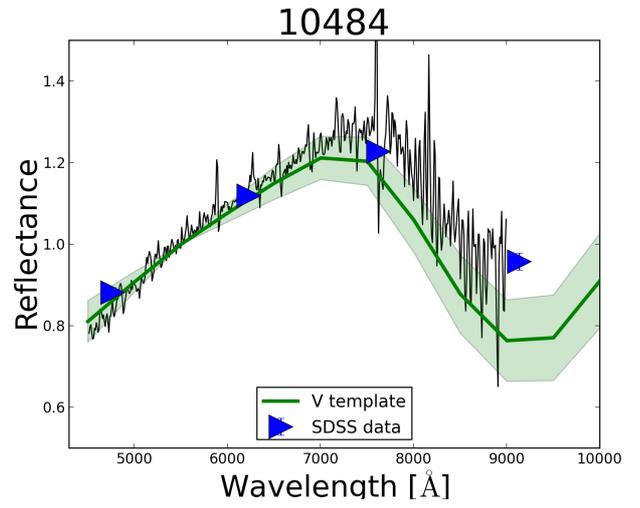}}
\caption{Same as in Fig. \ref{spec2} for asteroid (10484) Hecht.}
   \label{spec6}
 \end{figure}

\begin{figure}[htbp]
   \centering
\resizebox{\hsize}{!}{\includegraphics{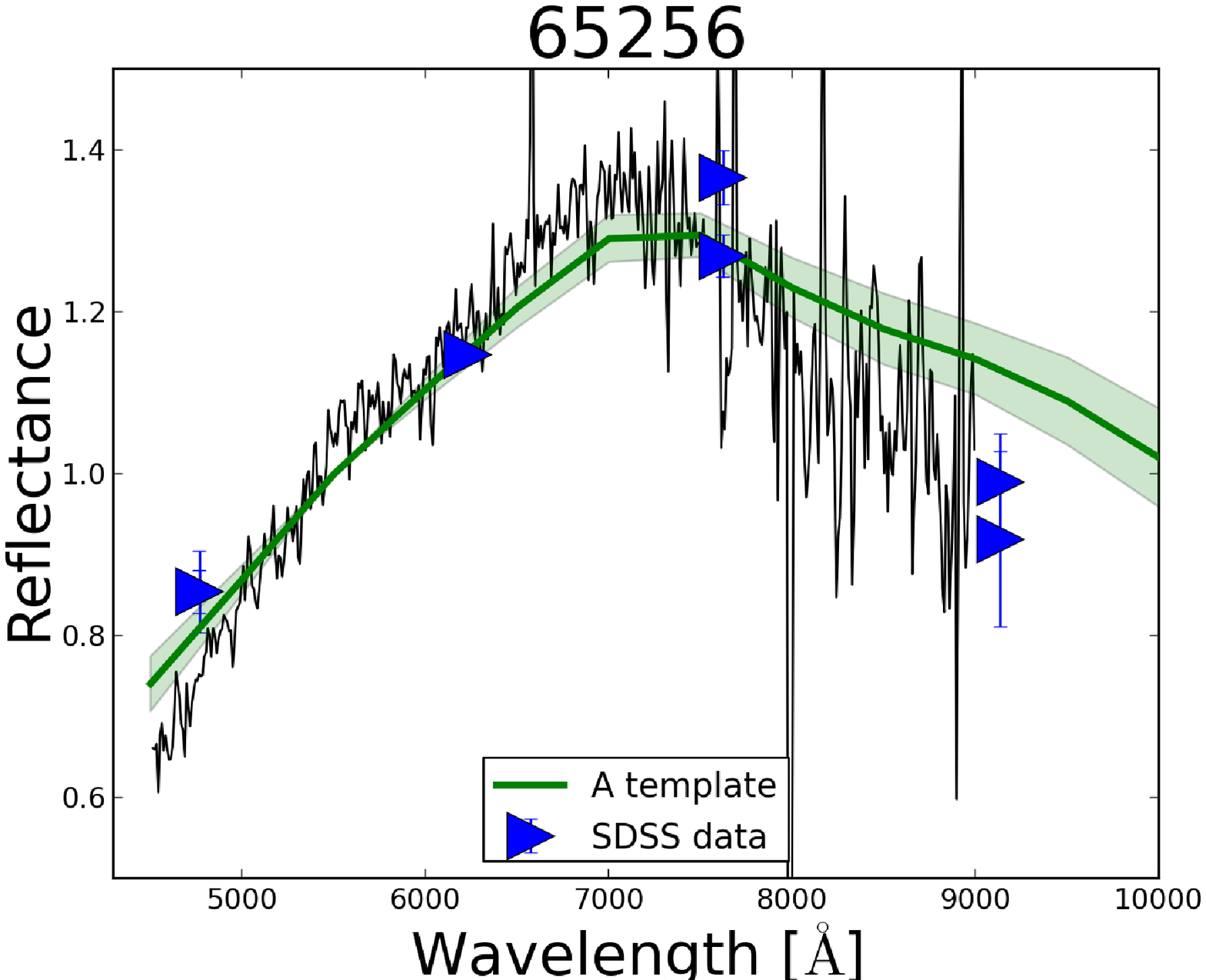}}
\caption{Spectra of asteroid (65256) 2002 FP34 obtained at the NOT. The triangles denote the reflectance values computed from the SDSS photometry. The thick line represents the taxonomical template for the A-type and the shaded area is the standard deviation of the template.}
   \label{spec7}
 \end{figure}

\begin{figure}[htbp]
   \centering
\resizebox{\hsize}{!}{\includegraphics{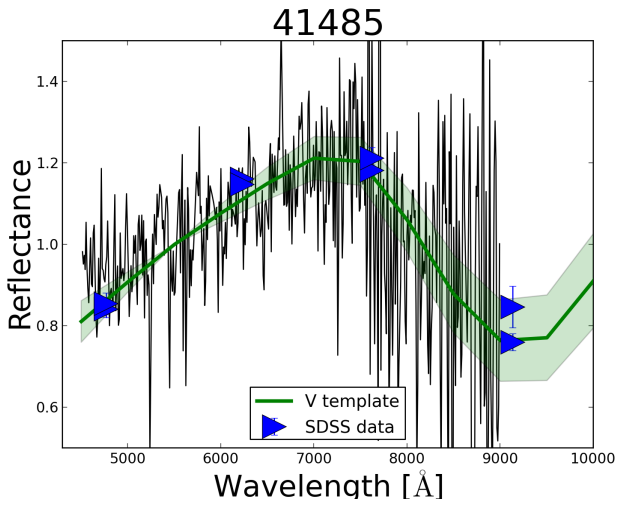}}
\caption{Same as in Fig. \ref{spec2} for asteroid (41485) 2000 QF51.}
   \label{spec8}
 \end{figure}

\begin{figure}[htbp]
   \centering
\resizebox{\hsize}{!}{\includegraphics{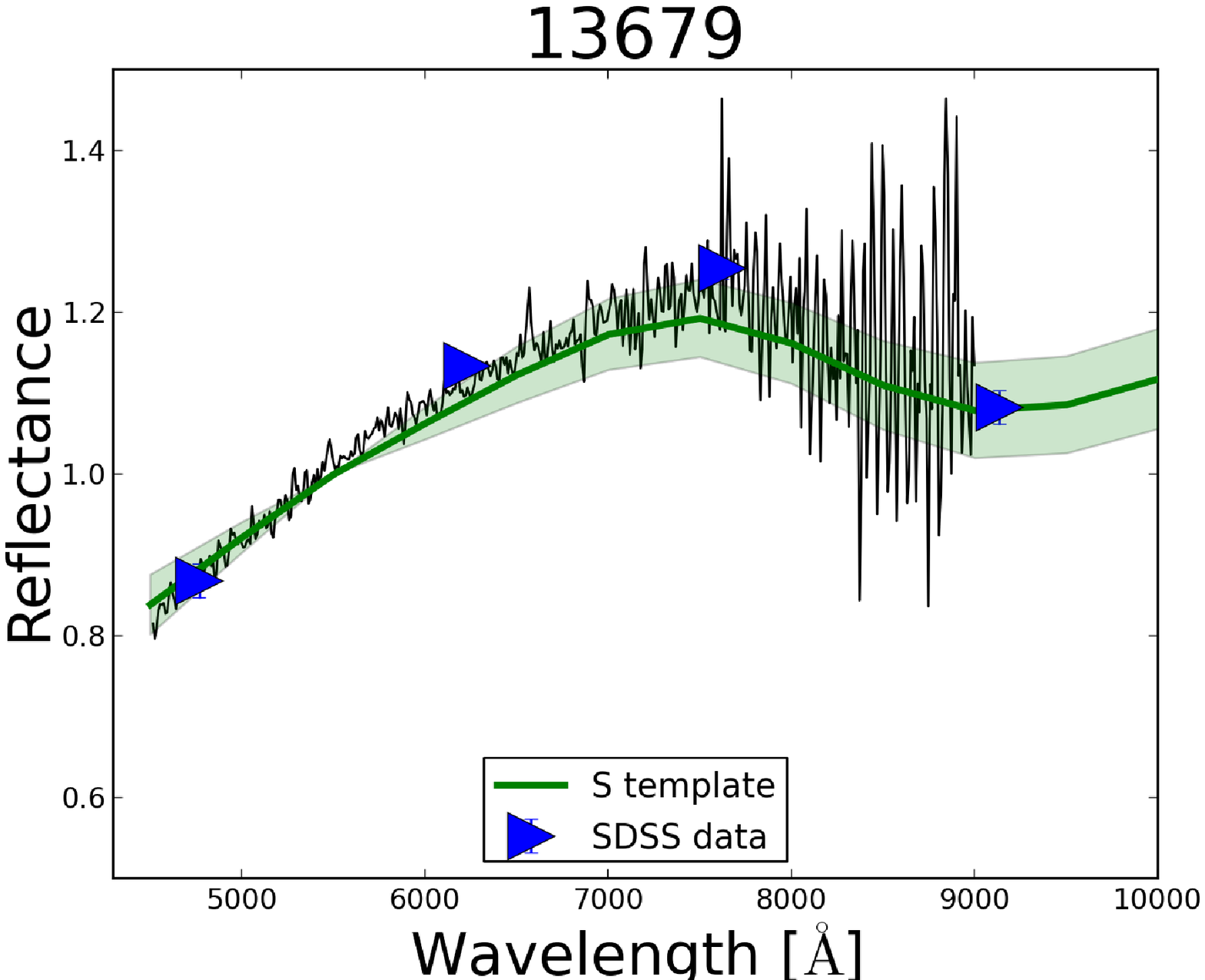}}
\caption{Spectra of asteroid (13679) Shinanogawa. The triangles denote the reflectance values computed from the SDSS photometry. The thick line represents the taxonomical template for the S-type and the shaded area is the standard deviation of the template.}
   \label{spec9}
 \end{figure}

\begin{figure}[htbp]
   \centering
\resizebox{\hsize}{!}{\includegraphics{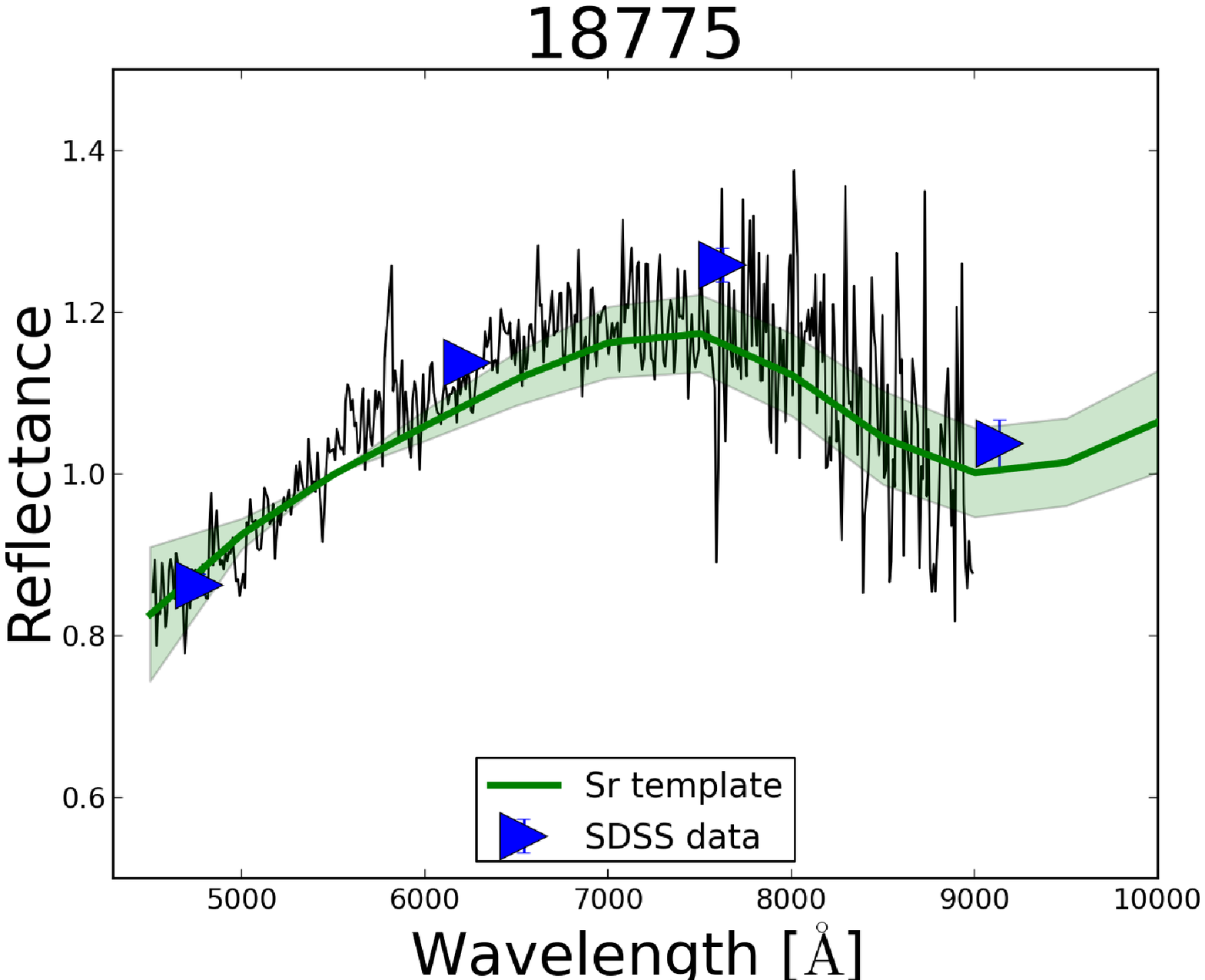}}
\caption{Spectra of asteroid (18775) Donaldeng. The triangles denote the reflectance values computed from the SDSS photometry. The thick line represents the taxonomical template for the Sr-type and the shaded area is the standard deviation of the template.}
   \label{spec10}
 \end{figure}

\begin{figure}[htbp]
   \centering
\resizebox{\hsize}{!}{\includegraphics{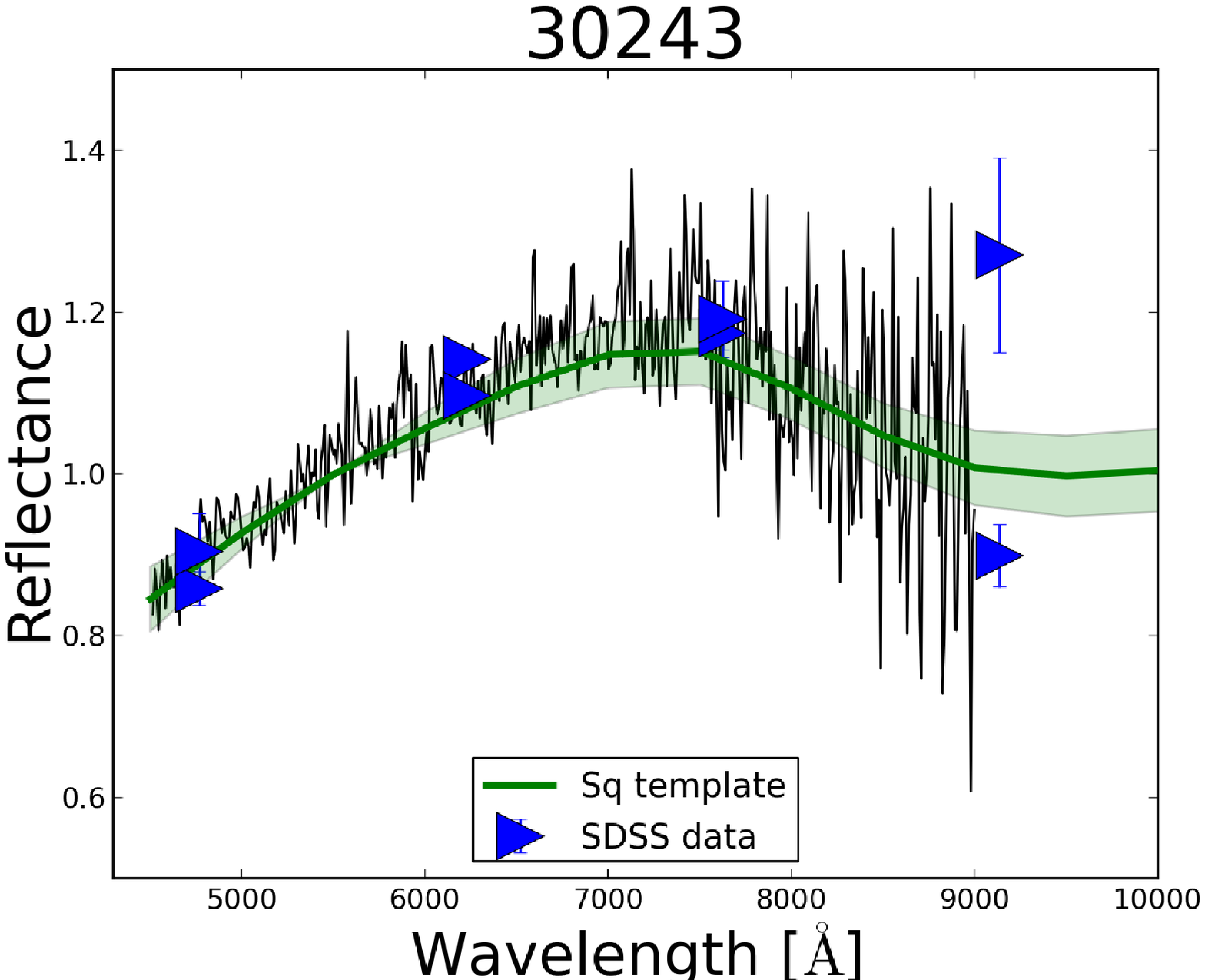}}
\caption{Spectra of asteroid (30243) 2000 HS9. The triangles denote the reflectance values computed from the SDSS photometry. The thick line represents the taxonomical template for the Sq-type and the shaded area is the standard deviation of the template.}
   \label{spec11}
 \end{figure}

\begin{figure}[htbp]
   \centering
\resizebox{\hsize}{!}{\includegraphics{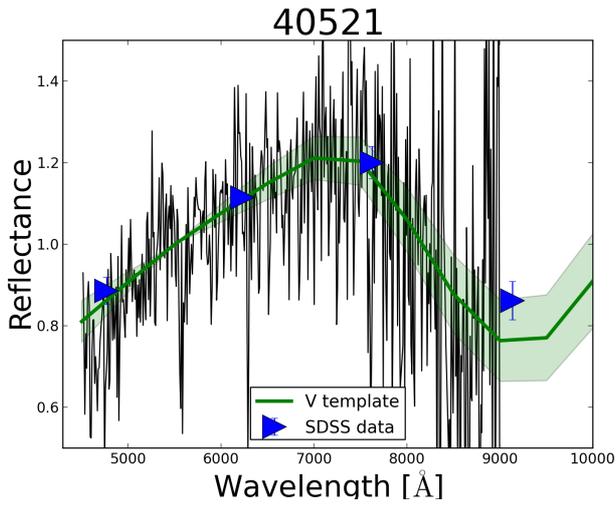}}
\caption{Same as in Fig. \ref{spec2} for asteroid (40521) 1999 RL95.}
   \label{spec12}
 \end{figure}

\begin{figure}[htbp]
   \centering
\resizebox{\hsize}{!}{\includegraphics{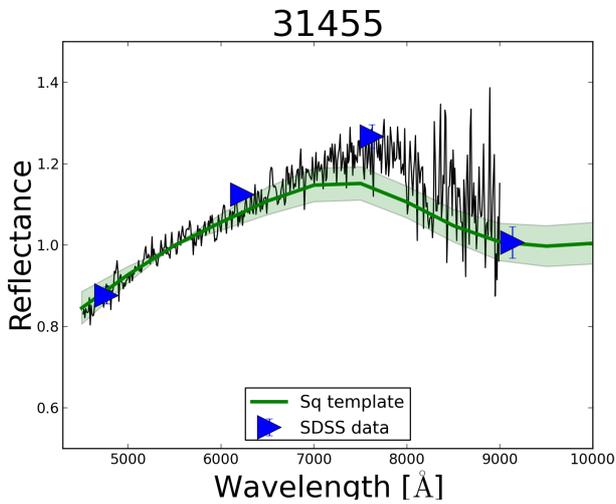}}
\caption{Same as in Fig. \ref{spec11} for asteroid 31455.}
   \label{spec13}
 \end{figure}

\begin{figure}[htbp]
   \centering
\resizebox{\hsize}{!}{\includegraphics{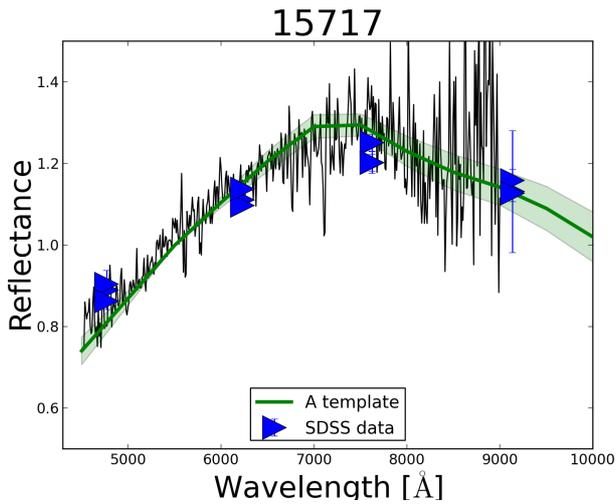}}
\caption{Same as in Fig. \ref{spec7} for asteroid (15717) 1990 BL1.}
   \label{spec14}
 \end{figure}

\begin{figure}[htbp]
   \centering
\resizebox{\hsize}{!}{\includegraphics{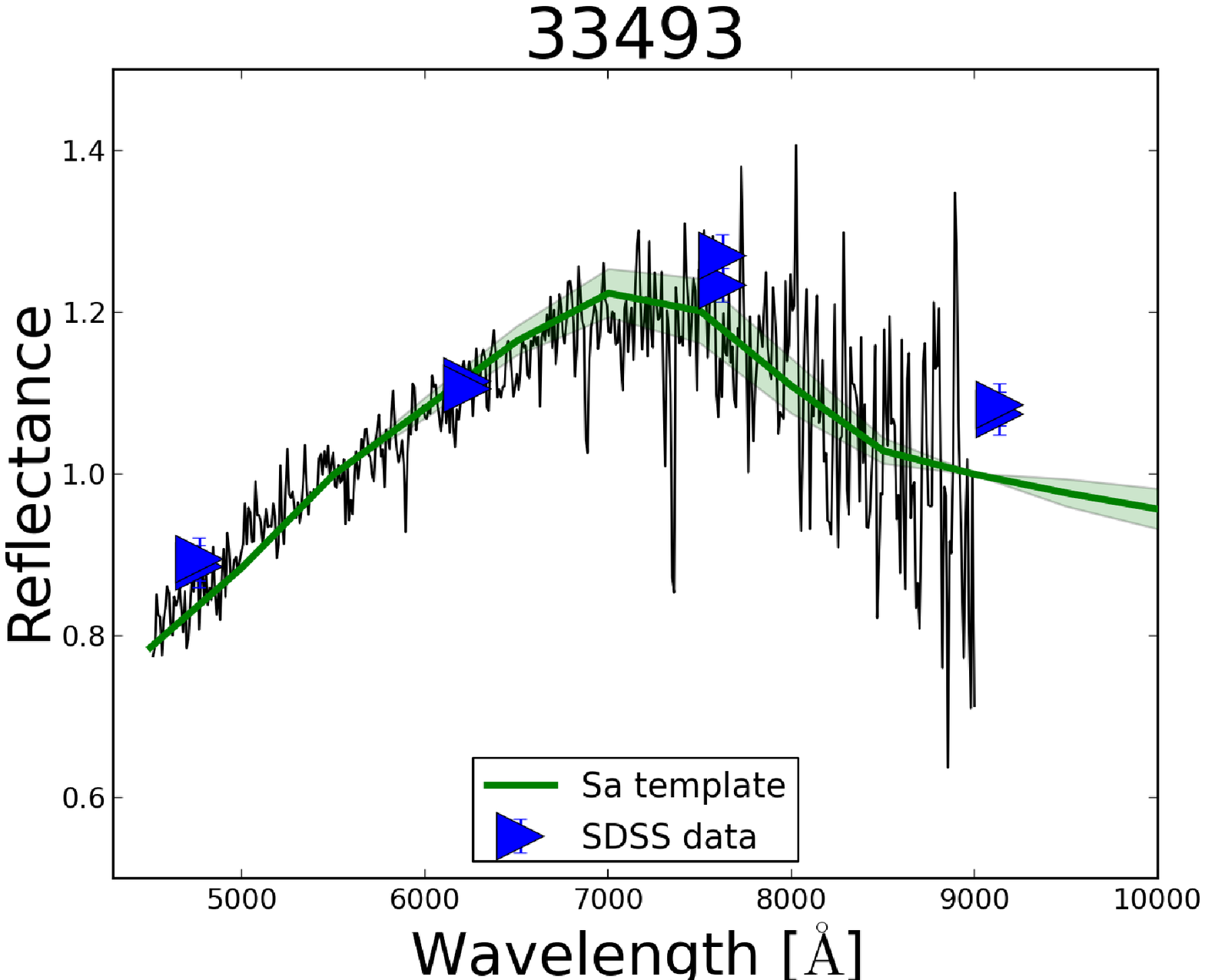}}
\caption{Spectra of asteroid 33493. The triangles denote the reflectance values computed from the SDSS photometry. The thick line represents the taxonomical template for the Sa-type and the shaded area is the standard deviation of the template.}
   \label{spec15}
 \end{figure}

\begin{figure}[htbp]
   \centering
\resizebox{\hsize}{!}{\includegraphics{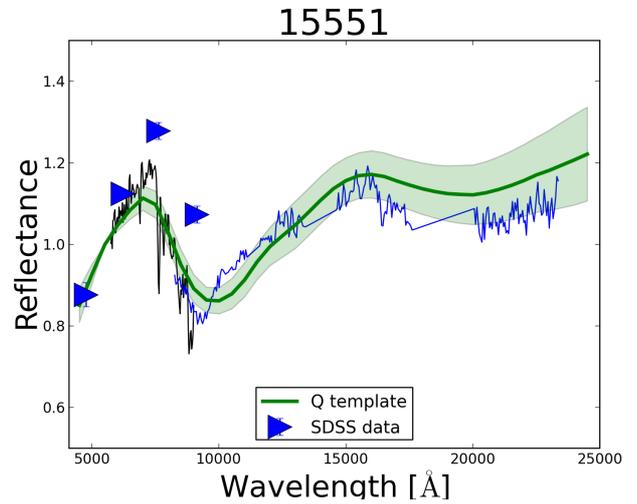}}
\caption{Spectra of asteroid (15551) Paddock obtained at the NOT and IRTF (courtesy of Francesca DeMeo) . The triangles denote the reflectance values computed from the SDSS photometry. The thick line represents the taxonomical template for the Q-type and the shaded area is the standard deviation of the template.}
   \label{spec16}
 \end{figure}

We present the obtained spectra in Figs. \ref{spec1} to \ref{spec16}. In Table \ref{obs} we present the observing circumstances along with the assigned taxonomic classification.
For most of the objects we observed 2 solar analogs. One before taking the science exposure, and second after the science exposure. The spectra of asteroids were divided by both solar analogs and then compared against each other. Asteroids were assigned preliminary taxonomical classes using the online classification tool \url{http://m4ast.imcce.fr/}.

\onecolumn

\begin{table}[htbp]
\caption{Observing circumstances and classification. Asteroids were classified using the M4AST tool \cite{Popescu2012}. The preliminary classification is determined by how closely the asteroid spectrum is fitted by the standard spectrum of each class using a curve matching algorithm. This involves fitting the spectrum with a polynomial curve and comparing this curve to the standard spectrum at the wavelength given in the taxonomy. The class producing the smallest standard deviation is then selected. The reliability is based on the number of points used for the comparison with the taxonomic templates. Please see \cite{Popescu2012} for details.}
   \label{obs}
   \centering
   \begin{tabular}{c c c c c c c c c c} 
   	\hline \hline
	Asteroid & Telescope & UT date & Phase  & V mag & Air  & Solar & preliminary & Std. dev. & Reliability \\ 
	 number   & 			& 	 	& angle &   & mass & analog & class & & of classification \\ \hline 
	
	15551 & NOT & 02-01-2013 & 19$^{\circ}$&  18.6 &1.2 & 51Peg  &  Q  &  0.042 & 85.3\% \\ 
	65256 & NOT & 07-05-2013 & 3$^{\circ}$& 18.8 & 1.5 &  107-998 & A & 0.062 (A) & 24.3\% \\
	1979 & NOT & 07-05-2013 & 25$^{\circ}$ & 18.2 & 1.2 & 102-1081, HD101364 & V & 0.279 & 26.8\% \\
	10484 & NOT & 07-05-2013 & 18$^{\circ}$ & 17.7 & 1.3 & HD101364, HD144873 & V & 0.132 & 29.2\% \\
	41485 & NOT & 07-05-2013 & 3$^{\circ}$ & 18.8 & 1.4 & 107-689, 107-998 & V & 0.077 & 21.9\%\\
	13679 & NOT & 07-05-2013 & 14$^{\circ}$ & 16.5 & 1.5 & 112-1333 & S-complex & 0.023 (S) &  24.3\% \\
	18775 & NOT & 07-05-2013 & 19 $^{\circ}$ & 18.5 & 1.4 & 107-998, HD144873 & S-complex & 0.05 (Sr) & 29.2\% \\ 
	
	30243 & NOT & 08-05-2013 & 5$^{\circ}$ & 18.2 & 1.6 & HD147284, 110-361 & S-complex & 0.04 (Sq) & 29.2 \% \\
	
	
	40521 & NOT & 11-05-2013 & 20$^{\circ}$ & 20.1 & 1.8 & HD147284, HD144873 & V & 0.102 & 29.2\%\\
	31455 & NOT & 11-05-2013 & 13$^{\circ}$  & 18.5 & 1.2 & HD144873, 102-1081 & S-complex & 0.030 (Sq) & 29.2 \% \\
	15717 & NOT & 11-05-2013 & 18$^{\circ}$ & 18.6 & 1.3 & 112-1333, HD147284 & A & 0.049 & 29.2\%\\
	33493 & NOT & 11-05-2013 & 18$^{\circ}$ & 18.0 & 1.5 & 112-1333 & S-complex & 0.020 (Sa) & 21.9 \% \\ 
	
	11699 & NOT & 03-07-2013 & 27$^{\circ}$ & 17.3 & 1.6 & HD148642, 102-1081 & V  & 0.026 & 24.3\% \\

	14419 & SALT & 21-03-2013 & 6$^{\circ}$ & 17.5 & 1.2 & 107-998, 107-684 & V & 0.060 & 19.5 \% \\
	
	16352 & SALT & 22-04-2013 & 21$^{\circ}$ & 17.6 & 1.2 & 107-684 & V & 0.018 & 21.9 \% \\
	
	11699 & SALT & 26-04-2013 &  10$^{\circ}$ &  15.8 & 1.2  & 102-1081 & - & - & - \\ 
	\hline
	
   \end{tabular}
\end{table}

\twocolumn

We targeted several inner Main Belt asteroids. Out of those (11699) 1998 FL$_{105}$, (14419) 1991 RK23, (41485) 2000 QF$_{51}$, and 16352 (1974 FF) are V types. 11699 was observed twice. First at the SALT and then because of a peculiar absorption around 0.65 microns the observations were repeated at the NOT. Second observation showed no absorption band around 0.65 microns. Therefore we conclude that the feature is most likely due to weather or instrumental causes. Asteroids (1979) Sakharov and (10484) Hecht are basaltic pairs from the inner Main Belt that we targeted for a different observing program (\cite{Polishook13}) during our run. Three other inner Main Belt asteroids (13679) Shinanogawa, (18775) Donaldeng, (39243) 2000 YU$_{76}$ are from the S-complex. Asteroid (40521) 1999 RL95 was already identified as basaltic by \cite{roig2008v}. Based on our spectra we also assign a V-type to this object. In the outer Main Belt we found several objects that are worth investigating in the NIR. Particularly asteroids (65256) 2002 FP$_{34}$, (15551) Paddock and (15717) 1990 BL1 have deep absorption bands. All those mid/outer Main Belt objects are especially interesting for investigating for the "missing dunite problem" \cite{burbine1996mantle}. If confirmed in NIR those object could contribute to the modest inventory of outer Main Belt objects of possible differentiated origin. For (15551) Paddock we have combined our data with NIR from the IRTF (courtesy of Francesca DeMeo) and classified it as an Q-type object. However it should be noted that the visible spectra resembles more a V-complex object and the NIR resembles S-complex object. It should also be mentioned that due to shorter wavelength coverage of grism 12, the normalization at 0.55 microns had to be done by extrapolation of the spectra to that region therefore leading to less reliable classification.

\section{Conclusions}
We have developed a new selection method. The method uses the SDSS photometric data, WISE albedos and G$_{12}$s to predict taxonomic type of asteroids. The method can be further improved when new objects are spectrally observed and classified. New classifications expand the training set and narrow the conditional probabilities. The efficiency of the method is lower compared to other methods, this is partly due to the fact that we allow noisy data to enter our classification. This procedure is purposeful. Even though the SDSS MOC has already been largely exploited we show that there is still room for new discoveries within this catalogue and merges with other large databases. For example targeting objects from the WISE database having high albedo can result in new discoveries of high albedo taxonomic types, such as for example the V type. 

We found several V type asteroids, most of them in the inner Main Belt. Three asteroids in the mid and outer Main Belt look especially interesting, namely: 65256, 15551, 15717. All of those show deep absorption bands near 1.0 micron and should be further investigated in the NIR. Additionally we also plan to target our strong V-type candidates in nearest opposition.

The distribution of V-type candidates shows a clustering around the Vesta family and scatter in the mid/outer Main Belt. Similarly to other studies we find that V-types candidates are not plentiful beyond $2.5$ AU. Given the increasing spectroscopic reach of the modern telescopes, we believe that dynamical, physical and geological paths should be explored in solving the missing dunite problem.

\begin{acknowledgements}
DO was supported by Polish National Science Center, grant number 2012/04/S/ST9/00022. TK was supported by the  Polish National Science Center, grant number N 203 403739.
Based on observations made with the Nordic Optical Telescope, operated by the Nordic Optical Telescope Scientific Association at the Observatorio del Roque de los Muchachos, La Palma, Spain, of the Instituto de Astrofisica de Canarias and observations made with the South African Large Telescope at the South African Astronomical Observatory, Sutherland, South Africa.

\end{acknowledgements}

\onecolumn

\begin{longtable}{llllll}
   \caption{V-type candidates in the mid and outer Main Belt. Objects already confirmed in previous studies are highlighted in bold.} 
 \label{candi} \\
 \hline\hline
ID & a [AU] & e & sin(i) & prob. & comments \\ 
\endfirsthead
\hline
71697 & 3.02 & 0.2 & 0.09 & 0.99 & \\
44447 & 2.57 & 0.16 & 0.24 & 0.62 & \\
58829 & 3.02 & 0.19 & 0.22 & 1.0 & \\
{\bf{21238}} & 2.54 & 0.14 & 0.19 & 1.0 & \\
83062 & 2.67 & 0.03 & 0.13 & 0.99 & \\
{\bf{10537}} & 2.85 & 0.1 & 0.11 & 0.6 & \\
34312 & 2.87 & 0.05 & 0.04 & 0.91 & \\
41243 & 2.99 & 0.08 & 0.19 & 1.0 \\
\hline
\end{longtable}

\Online
\begin{appendix}

\section{Target comparison} \label{targets}

\begin{longtable}{llllllll}

   \caption{Basaltic candidates targeted in other studies and their probability of being basaltic type as computed in this study. Probabilities have been rounded to two digits. In most of the cases probabilities based on different sets of observations agree with each other. When few datasets were available for object (for example asteroid observed few times by the SDSS) we have listed the highest probabilty of V-type or highest probabilty of any other class if no classification ended as V-type.
   } \label{comparison} \\
\hline \hline
Number & a & e & sin (i) & M08 & RH06 & AVAST &  Prob. \\ 
\hline
 \endfirsthead
\bf{28517}  & 2.29  &  0.09  &  0.14  &  True  &  True  &  True  &  V 0.82 \\
\bf{60669}  & 2.21  &  0.13  &  0.12  &  True  &  True  &  True  &  V 0.58 \\
\bf{24941}  & 2.48  &  0.12  &  0.09  &  True  &  False  &  True  &  V 0.53 \\
\bf{56570}  & 2.38  &  0.1  &  0.07  &  True  &  True  &  True  &  V 0.91 \\
\bf{46690 (S)}  & 2.24  &  0.1  &  0.07  &  True  &  True  &  True  &  V 0.72  \\
\bf{10537}  & 2.85  &  0.1  &  0.11  &  True  &  True  &  True  &  V 0.6 \\
55092  & 2.15  &  0.13  &  0.05  &  True  &  False  &  True  &  Sr 0.38 ( 0.03 ) \\
\bf{40521}  & 2.53  &  0.05  &  0.22  &  True  &  True  &  True  &  V 0.89 \\
\hline
\end{longtable}

\end{appendix}

\end{document}